# A Cloud-based Approach for Context Information Provisioning

Elarbi Badidi
Faculty of Information Technology
United Arab Emirates University
Al-Ain, United Arab Emirates
ebadidi@uaeu.ac.ae

Larbi Esmahi
School for Computing & Information Systems
Athabasca University, University Drive
Athabasca, Alberta, Canada
larbie@athabascau.ca

*Abstract*— As a result of the phenomenal proliferation of modern mobile Internet-enabled devices and the widespread utilization of wireless and cellular data networks, mobile users are increasingly requiring services tailored to their current context. High-level context information is typically obtained from context services that aggregate raw context information sensed by various sensors and mobile devices. Given the massive amount of sensed data, traditional context services are lacking the necessary resources to store and process these data, as well as to disseminate high-level context information to a variety of potential context consumers.

In this paper, we propose a novel framework for context information provisioning, which relies on deploying context services on the cloud and using context brokers to mediate between context consumers and context services using a publish/subscribe model. Moreover, we describe a multi-attributes decision algorithm for the selection of potential context services that can fulfill context consumers' requests for context information. The algorithm calculates the score of each context service, per context information type, based on the quality-of-service (QoS) and quality-of-context information (QoC) requirements expressed by the context consumer.

One of the benefits of the approach is that context providers can scale up and down, in terms of cloud resources they use, depending on current demand for context information. Besides, the selection algorithm allows ranking context services by matching their QoS and QoC offers against the QoS and QoC requirements of the context consumer.

## I. INTRODUCTION

The proliferation of wireless and cellular networks over the last few years has led to a remarkable rise in the number of users who are using a variety of modern mobile Internet-enabled devices --such as iPhones, iPads, and Android-based smartphones-- to consume online services. Mobile users are increasingly requiring services tailored to their context as they are on the move. Therefore, enterprise services should be context-aware to deal with the changing environment of the user. Several definitions of the notion of context have been provided in the literature. According to Dey [1], *"Context is any information that can be used to characterize the situation of an entity. An entity is a person, place, or object that is considered relevant to the interaction between a user and an application, including the user and applications themselves."*

According to this definition, the amount of information that can be categorized as context information is extremely wide. *Location*, *time*, *temperature*, *humidity*, *pressure*, and mobile user *activity* are the most widely used context indicators by applications. Specialized services, that we call *context services*, capture, store, analyze and aggregate data to provide high-level context information to consumer application services as needed. Context services and context consumers are often physically distributed. Besides, it is likely that these context sources provide the same context information but with different QoC [2][3]. The QoC concept is explained in Section 3. Context-awareness raises challenges like aggregation of context information in a structured format, discovery, and selection of appropriate context services for context delivery to context consumers.

To cope with the issues of context delivery and context service selection, we propose a novel framework for context provisioning, which is relying on using components called *context brokers*, and deploying context services on the cloud. Context brokers mediate between context consumers and context services using a publish/subscribe model. To the best of our knowledge there was no previous work on deploying context services on the cloud. We believe that our approach will take advantage of the power of the cloud in terms of elasticity, storage abundance, and scalability. Furthermore, we describe a multi-attributes algorithm for the selection of context services on the basis of the QoS and QoC they can offer. The algorithm takes into account the QoS and QoC requirements of context consumers for each context information to which they subscribe with the Context Broker.

The remainder of the paper is organized as follows. Section 2 describes related work on context-awareness and context information provisioning. Section 3 provides background information on the concepts of cloud services and quality-of-context. Section 4 presents an overview of our proposed framework, describes the interactions among the framework components and our proposed algorithm for the selection of context services in both a single cloud and multiple clouds. Section 5 discusses the challenges of the approach. Finally, Section 6 concludes the paper and describes future work.

II. RELATED WORK

Over the last two decades, context provisioning has been a particularly popular research topic, especially with the advent of smart mobile devices, the advances in sensing technology, and the proliferation of mobile applications. Many research works have proposed, designed, and implemented frameworks and middleware infrastructures for managing context information and providing users with context-aware services. Moreover, many surveys have been made in order to understand the features and shortcomings of existing systems [4][5][6].

With the emergence of service-oriented computing, numerous research works have investigated the design and the implementation of context services. A context service typically provides infrastructure support for collection, management, and dissemination of context information vis-à-vis a number of subjects. Subjects may be users, objects such as handheld devices and equipment, or the environment of users. The context service acquires context information from various context sources. For example, consider the "temperature" at the current location of the mobile user; this information may be obtained directly from the mobile device of the user. It can also be obtained from a local weather station. Alternatively, it may be obtained from weather TV channels providing weather information nation-wide.

Schmidt et al. designed and implemented a generic context service with a modular architecture that allows for context collection, discovery and monitoring [7]. This context service provides a Web service interface that allows its integration in heterogeneous environments. The implementation uses OWL to describe context information and SPARQL to query and monitor context information.

Lei et al. described the design issues and the implementation of a middleware infrastructure for context collection and dissemination [8]. They realize this middleware infrastructure as a context service. To allow for wide deployment of the context service, this work has addressed the following issues: extensibility of the context service architecture by supporting heterogeneous context sources, integrated support for privacy, and quality of context information support. Coronato et al. proposed a semantic context service that relies on semantic Web technologies to support smart offices [9]. It uses ontologies and rules to infer high-level context information, such as lighting and sound level, from low-level raw information acquired from context sources.

As it was described in the surveys mentioned earlier, many of the existing context-aware systems are suffering from the lack of scalability, extensibility, interoperability, and adoption difficulties. The originality of our approach lies in bringing context management and delivery to the cloud by deploying context services on the cloud. We believe that our approach will benefit from the power of the cloud in terms of scalability, elasticity, cloud storage abundance, and scaling up and down.

III. BACKGROUND

*A. Quality-of-Context*

Context information is characterized by some properties referred in literature as QoC indicators. Buchholz et al. [2] have defined the QoC as: *"Quality of Context (QoC) is any information that describes the quality of information that is used as context information. Thus, QoC refers to information and not to the process nor the hardware component that possibly provide the information."*

Buchholz et al. [2] and Sheikh et al. [3] have identified the following QoC indicators: *precision*, *freshness*, *temporal resolution*, *spatial resolution*, and *probability of correctness*.

Precision represents the granularity with which context information describes a real world situation. Freshness represents the time that elapses between the determination of context information and its delivery to a requester. Spatial resolution represents the precision with which the physical area, to which an instance of context information is applicable, is expressed. Temporal resolution is the period of time during which a single instance of context information is applicable. Probability of correctness represents the probability that a piece of context information is correct.

Several competing context services may provide the same context information [2]. Therefore, potential context consumers should be able to select context services on the basis of the QoC they can assure.

*B. Cloud services*

Cloud computing enables a service-provisioning model for computing services that relies on the Internet. This model typically involves the provisioning of dynamically scalable and virtualized services.

Applications or services offered by means of cloud computing are called *cloud services*. Typical examples of cloud services include office applications (word processing, spreadsheets, and presentations) that are traditionally found among desktop applications. Nearly, all large software corporations, such as Google, Microsoft, Amazon, IBM, and Oracle, are providing various kinds of cloud services. Besides, many small businesses have launched their own Web-based services, mainly to take advantage of the collaborative nature of cloud services.

The user of a cloud service has access to the service through a Web interface or via an API. Once started, the cloud service application acts as if it is a normal desktop application. The difference is that working documents are on the cloud servers.

Cloud services models are:

- *Infrastructure-as-a-Service (IaaS)*: With IaaS, organizations rent computing resources and storage space and access them through a private network or across the Internet.

- *Platform-as-a-Service (PaaS)*: With PaaS, organizations can develop their business applications in a cloud environment by using software tools supported by their cloud provider. Maintenance and management of the cloud infrastructure including severs and operating system is the responsibility of the cloud provider.

- *Software-as-a-Service (SaaS)*: With SaaS, the cloud service application runs on the cloud provider servers and users access the service through a Web interface or via an API.

IV. A FRAMEWORK FOR CLOUD-BASED CONTEXT PROVISIONING

In every business with a delivery/consumption model, brokers emerge to mediate between consumers and providers. This could be the case for context delivery. Context brokers may, then, be used to decouple context consumers from context services. Our interest in using brokers is motivated by the fact that they have been used for a while in Service Oriented Architecture (SOA) to mediate between services providers, service consumers, and partners. They have also been extensively used in multimedia systems and in mobile computing systems to deal mainly with the issue of QoS management.

Fig. 1 depicts our framework for context information provisioning. The main components of the framework are: *Context-aware Web services (context consumers)*, *Context Brokers*, and *Cloud-based Context Services*. Multiple context brokers may be deployed, one for each local domain for instance. A discovery service will allow context-aware consumers to bind to the right context broker.

*A. Context Brokers*

A *context broker* is a mediator service that decouples context consumers from context services. It is in charge of handling subscriptions of context consumers in which they express their interest to receive context information, and registration of context services. Context services may then publish their newly acquired context information to the context broker, which notifies context consumers about that newly acquired context information. Context brokers can also be deployed on the cloud. Fig. 2 illustrates our topic-based publish-subscribe system in which context services are the publishers and the CAWSs are the subscribers.

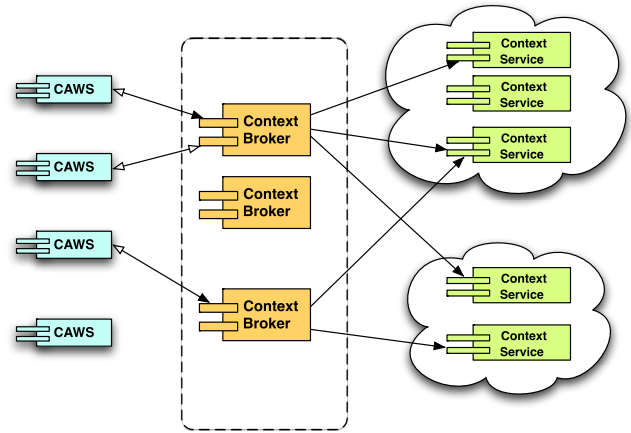

**CAWS: Context-Aware Web Service**

Figure 1. Framework for Cloud-based Context Provisioning

Context information -- such as *location*, *temperature*, and *user activity* -- represents the topics of the system. The Publish/subscribe messaging model is a one-to-many pattern of asynchronous message distribution based on registration of interest. In this model, publishers associate the name of a topic to each message ("publish") rather than addressing it directly to subscribers. Then, the message system sends the message to all eligible recipients that expressed their interest in receiving messages on that topic ("subscribe"). As opposed to point-to-point messaging systems, such as message queuing, the publish/subscribe model of asynchronous communication is a far more scalable architecture. This is because the source of the information has only to concern itself with creating the information, and can leave the task of servicing potential recipients to the messaging system. It is a loosely coupled architecture in which senders often do not need to know who their potential subscribers are, and the subscribers do not need to know who generates the information.

In addition to this publish/subscribe model for provisioning context information, a context broker implements a regular on-demand request/response model, in which it requests up-to-date context information from context services once a context consumer requires information for a given topic. Therefore, a context broker may either pull context information from context services or let context services push updated context information.

Context services, typically residing in different clouds, deliver context information to context consumers with various quality-of-context and quality-of-service (QoS). Therefore, the Context Broker is in charge of selecting appropriate context services to deliver context information to which a context consumer has subscribed. Context information may be delivered to the same consumer by several context services. Each one may deliver a piece of context information (a topic) that the consumer requires to adapt its behavior to the current context of a user. In Sub-section 4.5, we describe a selection algorithm that allows ranking context services with regard to the QoC and the topics required by a context consumer.

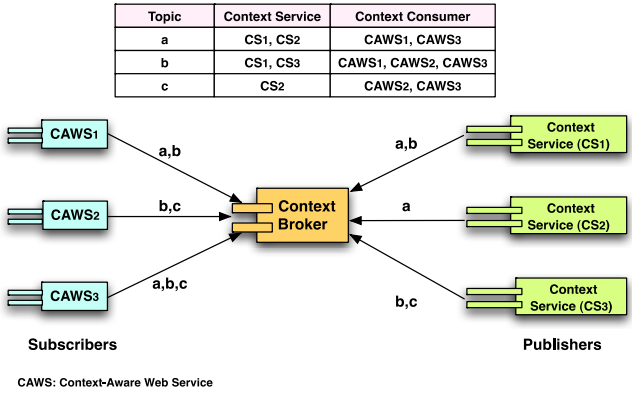

Figure 2. Topic-based publish/subscribe system

*B. Context Consumers*

In our framework, context-aware Web services (CAWS) are the consumers of context information obtained from the cloud-based context services. A CAWS is a Web service that can understand situational context and can adapt its behavior according the changing circumstances as context data may change rapidly. It produces dynamic results according to the 5 WH questions: who, where, when, what, and why it was invoked. A CAWS can be responsive to various situational circumstances, such as:

- The identity of the client who invoked the service, whether it is a person, or another Web service.

- The location of the client.

- The time at which the client invokes the service.

- The activity that the client is carrying out at the time it invokes the service.

- The preferences that the client may have defined prior to invoking the service.

- The security and privacy policies associated with the client of this service.

- The device (laptop, PDA, smartphone, etc.) that the client is using to invoke the service.

*C. Cloud-based Context Services*

As we have mentioned earlier in the related work section, high-level context information is typically obtained from context services that aggregate raw context information sensed by sensors and mobile devices. Given the massive amount of context data processed and stored by context services and the wide acceptance of the cloud computing technology, context providers now can leverage their services by deploying them on the cloud.

Fig. 3 depicts the process of context acquisition and the deployment of context services on the cloud to provide high-level context information to context consumers. Raw context data sensed by various devices and sensors is processed, aggregated by *Context Aggregator* components in a structured format, and then uploaded to the cloud-based context services.

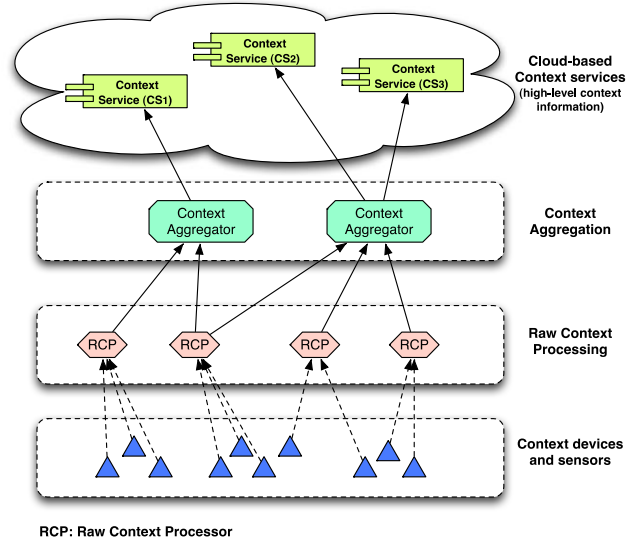

Figure 3. Deployment of high-level context information on the cloud

One of the underlying advantages of the deployment of context services in the cloud is the economy of scale. By making the most of the cloud infrastructure provided by a cloud vendor, a context provider can offer better, cheaper, and more reliable services than is possible within its premises. The context service can utilize the full processing and storage resources of the cloud infrastructure if needed. Another advantage is scalability in terms of computing resources. Context providers can scale up when additional resources are required as a result of a rise in the demand for context information. Conversely, they can scale down when the demand for context information is decreasing. Another benefit of the approach is to enable context-aware application services to acquire their required context information on a pay-as-you-go basis and to select cloud-based context services on the basis of the price they have to pay and other criteria, such as the QoC they can get. Furthermore, context-aware applications can obtain context information from cloud-based context services without having to be involved in context management. The net benefit for consumers and mobile users, in particular, is the ability to receive better services tailored to their current context.

The SaaS model is the most appropriate model for cloud-based context provisioning. Indeed, SaaS is seen as the trend of the future and the most common form of cloud service development. With SaaS, software is deployed over the Internet and delivered to thousands of customers. Using this model, the context service provider may license its service to customers through a subscription or a pay-as-you-go model. The service is then accessible using an API.

*D. Interfaces and Interaction model*

In this section, we describe the interactions among the components of the framework and do consider only the case of a single context broker. The model can be easily extended to consider several context brokers. Fig. 4 shows a simplified class diagram of the framework components, and Fig. 5 depicts the interactions among them.

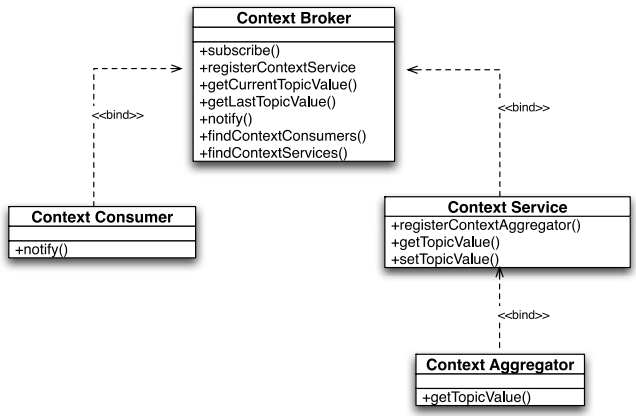

Figure 4- Class diagram of the framework components.

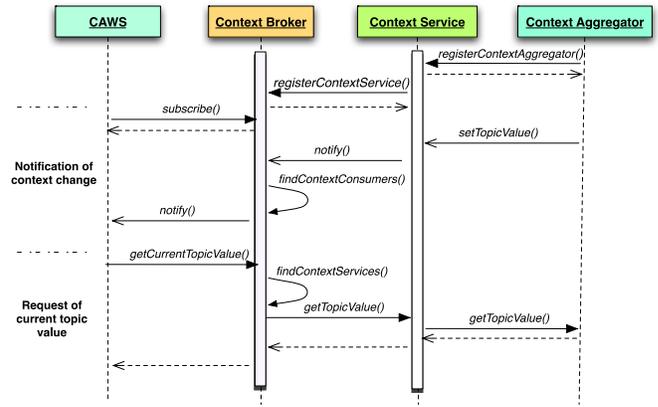

Figure 5- Diagram of interactions among the framework components

The context broker acts as an intermediary between publishers (context services) and subscribers (context consumers) on a collection of topics (context information).

A context consumer invokes the *subscribe()* method of the context broker to register its interest to receive updates on some topics, such as location, and temperature. If the processing of *subscribe()* is successful, the context broker returns a subscription ID to the context consumer.

Similarly, a context service invokes *registerContextService()* of the context broker to register its interest to publish some types of context information through the context broker. If the processing of that method is successful, the context broker returns a registration ID to the context service.

The context broker receives notifications of context change through its *notify()* method that a context service invokes. It, then, notifies a context consumer about context change by invoking its *notify()* method. Furthermore, a context consumer may request the current value for a given topic by invoking *getCurrentTopicValue()* of the context broker. The broker forwards the request to context services that are providing that topic requested by the context consumer. A newly-subscribed context consumer can invoke *getLastTopicValue()* in order to get the last value of a given topic that other consumers have already received.

The context broker has also two additional methods *findContextConsumers()* and *findContextServices()* that are self-invoked. The former is invoked to get the list of context consumers that have subscribed to a given topic once a notification of context change has been received for that topic. The latest is invoked to get the list of context services that are publishing the topic requested by a context consumer that has invoked *getCurrentTopicValue()*.

A context aggregator can register at a context service by specifying what topics it is an aggregator for. Once registered, a context aggregator can submit the current value for a given topic by invoking the *setTopicValue()* method at the context service. When the topic value is changed in the context service, the *notify()* method at the context broker is triggered to notify all subscribers of that topic.

*E. A Multi-attributes Algorithm for Context Services Selection*

As we have stated earlier, the *Context Broker* is in charge of selecting suitable context services to deliver context information to which context consumer (CAWS) subscribed. Context information may be delivered to the same context consumer by several context services. Each one may deliver a piece of context information (a topic) that the context consumer requires to adapt its behavior to the current context of a user. Thus, the selection has to be done per topic. In this subsection, we describe our proposed algorithm for context services selection. The algorithm allows ranking context services with regard to the QoC and the QoS required by a context consumer. We first describe how the algorithm works in the case of a single cloud; then, we extend the algorithm to the case of multiple clouds as depicted by Fig. 1.

*1) Single Cloud-based Service Selection*

As numerous potential context services, within the cloud, can deliver the context information required by a consumer, it is indispensable to consider only potential context services that can satisfy both the QoC and the QoS required by the context consumer.

Let $T = \{t_1, t_2, \ldots, t_C\}$ be the list of context information (topics) to which a context consumer has subscribed by showing its interest in receiving such context information. Let $CS = \{CS_1, CS_2, \ldots, CS_K\}$ be the list of context services in the cloud *that* have subscribed with the *Context Broker*. Two context services may provide different context information; each one specializes in offering particular context information. One service, for example, may offer *location* information while another service may offer only *temperature* information, and a third one may offer both of them.

These services typically provide context information with different QoC and QoS. We assume that QoC and QoS indicators are in normalized form with values between 0 and 1. A value of 1 means highest quality and 0 means lowest quality. For example for the *freshness* quality indicator, 1 means that context sources have sensed the information in the last minute, and 0 means that they have sensed it in the last 10 minutes. QoS indicators may concern for instance parameters such as *availability*, *response- time*, *reputation*, and *cost of service*.

When subscribing to context information, a context consumer specifies the min values of the normalized QoC and QoS indicators that he can tolerate. For instance, the context consumer may subscribe to the *location* information may require a min value of 80% for the *freshness* indicator, 93% for the *probability of correctness* indicator. He may require also 98% for the *Availability* QoS indicator. Let $P = \{P_1, P_2, ..., P_m\}$ be the list of QoC indicators (parameters) considered in the system. Let $Q = \{Q_1, Q_2, ..., Q_n\}$ be the list of QoS indicators considered in the system.

The minimum QoC requirements that the context consumer tolerates for a given context information (topic) $t_j$, with $1 \leq j \leq C$, are expressed by the following vector:

$$M_j = \{min_{1,j}, min_{2,j}, ..., min_{m,j}\}$$

$0 \leq min_{i,j} \leq 1$, with $1 \leq j \leq C$ and $m$ is the cardinality of $P$.

Therefore, the whole quality-of-context requirements of the context consumer for all its subscribed topics and all QoC indicators considered in the system can be expressed by the following matrix:

$$T = \begin{array}{c} \\ t_1 \\ t_2 \\ ... \\ ... \\ t_C \end{array} \begin{array}{cccc} P_1 & P_2 ........ & & P_m \\ \left[ \begin{matrix} min_{1,1} & min_{2,1} \cdots & \cdots & min_{m,1} \\ & \vdots & & \vdots \\ min_{1,2} & min_{2,2} \cdots & \cdots & min_{m,2} \\ & \vdots & & \\ min_{1,C} & min_{2,C} \cdots & \cdots & min_{m,C} \end{matrix} \right] \end{array}$$

The minimum QoS that the context consumer tolerates concern all topics, are expressed by the following vector:

$S = \{min_1, min_2, ..., min_n\}$ . $min_k$ represents the minimal value that the context consumer is willing to accept for the QoS parameter $Q_k$, for $1 \leq k \leq n$

A zero value in any QoC or QoS parameter means that the user has not specified any constraint on that parameter.

The goal of the selection algorithm is to find for each topic $t_j$, to which the context consumer subscribed, a suitable context service from the set $CS$ that can satisfy the minimum quality requirements of the context consumer.

The QoC offer of a context service $CS_r$ is expressed by the following matrix:

$$T_r = \begin{array}{c} \\ t_1 \\ t_2 \\ ... \\ ... \\ t_C \end{array} \begin{array}{cccc} P_1 & P_2 ........ ... & & P_m \\ \left[ \begin{matrix} q_{1,1}^r & q_{2,1}^r \cdots & \cdots & q_{m,1}^r \\ & \vdots & & \vdots \\ q_{1,2}^r & q_{2,2}^r \cdots & \cdots & q_{m,2}^r \\ & \vdots & & \\ q_{1,C}^r & q_{2,C}^r \cdots & \cdots & q_{m,C}^r \end{matrix} \right] \end{array}$$

The QoS offer of $CS_r$ is expressed by the following vector:

$Q^r = [q_1^r, q_2^r, ..., q_n^r]$, Where $q_k^r$ is the offer of $CS_r$ for the QoS indicator $Q_k$; $1 \leq k \leq n$.

The quality-of-service requirements of the context consumer are independent from the topics.

$CS_r$ is suitable for a topic $t_j$ if the following condition is satified:

$$0 \leq min_{i,j} \leq q_{i,j}^r \leq 1 \text{ for } 1 \leq i \leq m \text{ and } 1 \leq j \leq C$$

$$\text{and } min_k \leq q_k^r \quad \text{for } 1 \leq k \leq n \quad (1)$$

In other words, $CS_r$ is suitable for provisioning topic $t_j$ if the minimum quality-of-context requirements as well as the minimum quality-of-service requirements are satisfied.

In the following, we will consider in the selection process only context servers that meet the minimum QoS requirements of the context consumer.

The context consumer may set relative weights for the QoC indicators. He may even set weights for each topic to which it subscribed. For example, for the *location* topic, more weight may be given to the *spatial resolution* indicator than to the *probability of correctness* indicator. For the *time of the day* topic, more weight may be given, for example, to the *precision* indicator than to the other QoC indicators. Therefore, the weight matrix is given by:

$$W = \begin{array}{c} \\ t_1 \\ \\ t_2 \\ ... \\ ... \\ t_C \end{array} \begin{array}{cccc} P_1 & P_2 ........ & & P_m \\ \left[ \begin{matrix} w_{1,1} & w_{2,1} \cdots & \cdots & w_{m,1} \\ & \vdots & & \vdots \\ w_{1,2} & w_{2,2} \cdots & \cdots & w_{m,2} \\ & \vdots & & \\ w_{1,C} & w_{2,C} \cdots & \cdots & w_{m,C} \end{matrix} \right] \end{array}$$

The score of a given QoC indicator $P_i$ for a given topic $t_j$ by the $CS_r$ offer is:

$$s_{i,j}^r = w_{i,j} \times q_{i,j}^r$$

$$\text{for } 1 \leq i \leq m \text{ and } 1 \leq j \leq C \quad (2)$$

Therefore, the score matrix $S_r$ of the $CS_r$ offer, for all QoC indicators and all topics is:

$$S_r = \begin{array}{c} \\ t_1 \\ \\ t_2 \\ ... \\ ... \\ t_C \end{array} \begin{array}{cccc} P_1 & P_2 ........ .. & & P_m \\ \left[ \begin{matrix} s_{1,1}^r & s_{2,1}^r \cdots & \cdots & s_{m,1}^r \\ & \vdots & & \vdots \\ s_{2,1}^r & s_{2,2}^r \cdots & \cdots & s_{m,2}^r \\ & \vdots & & \\ s_{1,C}^r & s_{2,C}^r \cdots & \cdots & s_{m,C}^r \end{matrix} \right] \end{array}$$

Given the weight matrix and the minimum QoC requirements matrix, the minimum score matrix is:

$$S_{min} = \begin{array}{c} \\ t_1 \\ \\ t_2 \\ ... \\ ... \\ t_C \end{array} \begin{array}{cccc} P_1 & P_2 ........ & & P_m \\ \left[ \begin{matrix} l_{1,1} & l_{2,1} \cdots & \cdots & l_{m,1} \\ & \vdots & & \vdots \\ l_{1,2} & l_{2,2} \cdots & \cdots & l_{m,2} \\ & \vdots & & \\ l_{1,C} & l_{2,C} \cdots & \cdots & l_{m,C} \end{matrix} \right] \end{array}$$

Where $l_{i,j} = w_{i,j} \times min_{i,j}$

for $1 \leq i \leq m$ and $1 \leq j \leq C$

The difference matrix, $S_r - S_{min}$, shows whether $CS_r$ may satisfy or not all QoC requirements for all topics to which the context consumer has subscribed to. A value that is less than zero in this matrix means that $CS_r$ cannot satisfy the QoC requirement for the associated topic and QoC indicator.

Therefore, we have to reason per topic, and consider only context services that can meet the QoC requirement for that topic.

The score per topic $t_j$ for a potential context service $CS_r$ offer is:

$$score_j^r = \sum_{i=1}^m score_{i,j}^r. \quad (3)$$

The score of $CS_r$ for all topics can be expressed by the following vector:

$$V_r = \begin{bmatrix} score_1^r \\ score_2^r \\ \dots \\ \dots \\ score_C^r \end{bmatrix}$$

Considering the scores of all the potential context services, we get the following decision matrix:

|       | $CS_1$ | $CS_2$ | … | $CS_K$ | Max score | Selected CS |
|-------|--------|--------|---|--------|-----------|-------------|
| $t_1$ | $score_1^1$ | $score_1^2$ | … | $score_1^K$ | … | … |
| $t_2$ | $score_2^1$ | $score_2^2$ | … | $score_2^K$ | … | … |
| …     | … | … | … | … | … | … |
| $t_C$ | $score_C^1$ | $score_C^2$ | … | $score_C^K$ | … | … |

A score in the decision matrix is zero if the context service cannot meet the QoC requirements for a given topic.

The maximum score value of each row $j$ corresponds to the best QoC offer that can fulfill the QoS and QoC requirements of the context consumer for the topic $t_j$.

The most suitable context service for topic $t_j$, that we call here $Selected_j$, will be the one that maximizes the above score, that is:

$$Selected_j \leftarrow \max_{1 \leq r \leq K} (score_j^r). \quad (4)$$

If no context service satisfies the context consumer QoS and QoC requirements for a given topic, then the *Context Broker* may ask the context consumer to lower its QoC expectations.

The steps of the algorithm are summarized in Fig. 6.

*2) Multiple Clouds-based Service Selection*

The previous subsection describes how the ranking and selection of context services is achieved within a single cloud. In order to find out the most suitable context services, for each topic, within multiple clouds, the context broker selects potential context services in each cloud according to the algorithm described in the previous sub-section. Selected context services from the clouds are then ranked to find out the best context services per topic, which maximizes the score expressed by equation (3).

---

***Step-1***: *Construct the matrix **T** of minimum QoC requirements of the context consumer for all the topics it subsribes to, and the vector **S** of minimum QoS requirements the context consumer can tolerate. We assume that all values of the matrix and the vector are normalized to be in the range [0,1].*

***Step-2***: *Construct the weight matrix **W** set by the context consumer for each topic and for each QoC indicator, then the minimum score matrix $S_{min}$.*

***Step-3***: *For each Context service $CS_r$ registered with the Context Broker,*
a) *Construct the normalized matrix $T_r$ of the QoC offers of $CS_r$ for all current topics to which the context consumer has subscribed to, and the normalized vector $Q_r$ of the QoS offer of $CS_r$.*
b) *Calculate the score matrix $S_r$ that represents the score between the QoC offer of $CS_r$ and the context consumer QoC requirements for each quality indicator considered in the system and for each topic.*
c) *Calculate the difference matrix, $S_r - S_{min}$. If a value of this matrix is less than zero, then it means that $CS_r$ cannot satisfy the QoC requirements of the context consumer for the associated topic and the associate QoC indicator. Only rows with positive values will be considered in the next steps.*
d) *Calculate the score vector $V_r$ using equation (3). Note that rows with negative values in the difference matrix will have a score 0 in the score vector.*

***Step-4***: *Create the decision matrix, and fill out the maximum score for each topic and the CS providing that score.*

---

Figure 6. QoC-based Context Service Selection Algorithm

V. CHALLENGES OF THE APPROACH

In conjunction with the benefits provided by the cloud, deploying context services to the cloud raises numerous issues for context providers to consider, including possible interoperability, security, and performance concerns.

The interaction model described in the previous section provides the basis for the development of a context service API that will be used by both context brokers and context consumers to interact with context services. Heterogeneity of the APIs offered by various context services will be one of the challenges of the approach, especially if they are residing on different clouds. Context brokers should, then, be able to interoperate with all these heterogeneous context services.

Security is a significant concern with any SaaS application on the cloud. Care must be taken when designing and implementing a security solution for a cloud-based context-service to keep it as simple and efficient as possible. For instance, the context service may have to be integrated with an identity management service. In this scenario, each customer of the context service has an identity account, which is used to authenticate the customer and track all its requests for service.

Performance monitoring, billing, managing customers' expectations are also significant concerns among others that a context service provider has to handle. The context provider must ensure that its context service is highly available and that its customers can access it. One outage or crash of the service can affect all its customers. Now, there is a general trend toward implementing a Service Level Agreement (SLA) between providers of cloud services and customers, even though that most SaaS vendors do not provide them at present.

Another concern, which is not linked to the cloud, but that should be handled by context brokers and consumers is the heterogeneity in the representation and modeling of context information by each context service. Bettini et al. [10] provide a survey in which they describe and compare current context modeling and reasoning techniques. Strang et al. [11] provide another similar survey. Modeling approaches mainly include key-values models, graphical models, object-oriented models, markup scheme models, logic-based models, and ontology-based models. With this heterogeneity in context information models, context brokers should provide a common ontology-based context information model and the mappings from the various models to this common model.

## VI. CONCLUSION AND FUTURE WORK

High-level context information is typically obtained from context services that aggregate raw context information sensed by sensors and mobile devices. Given the enormous amount of context data processed and stored by context services and the wide acceptance of the cloud computing technology, context providers now can leverage their services by deploying them on the cloud.

In this paper, we have presented our proposed framework for cloud-based context provisioning. The framework relies on context brokers for context information dissemination using a publish/subscribe model. Context services, deployed on the cloud, can scale up and down, in terms of cloud resources they use, according to the demand for context information. We have described a preliminary model of interactions, among the components of the framework, and that could be the basis for a context service API. As a future work, we first intend to investigate further on a common ontology-based model for context information representation that can be used by context brokers; and then, describe the mappings from the various context representation models described in the literature to that common model. We also intend to implement a prototype of the framework by considering some real scenarios for context provisioning, and implementing a context broker and few similar cloud-based context services using open-source software tools.

## REFERENCES

[1] A.K. Dey, "Understanding and Using Context," Journal of Pervasive and Ubiquitous Computing, vol. 5(1), pp. 4–7, 2001.
[2] T. Buchholz, A. Kpper, M. Schiffers, "Quality of context: What it is and why we need it?," In Proc. of the 10th International Workshop of the HP OpenView University association (HPOVUA)*, 2003*.
[3] K. Sheikh, M. Wegdam, and M. Van Sinderen, "Quality-of-Context and its use for Protecting Privacy in Context Aware Systems," Journal of Software, vol. 3(3) pp. 83-93, March 2008.
[4] M.Baldauf, S. Dustdar, and F. Rosenberg, "A survey on context-aware systems," International Journal of Ad Hoc and Ubiquitous Computing, vol. 2 (4), pp. 263-277, 2007.
[5] K. Henricksen, J. Indulska, T. McFadden, and S. Balasubramaniam, "Middleware for Distributed Context-Aware Systems," OTM Confederated International Conferences, pp. 846-863, Springer-Verlag, 2005.
[6] H.L. Truong, and S. Dustdar, "A Survey on Context-aware Web Service Systems," International Journal of Web Information Systems, vol. 5(1), pp.5-31, Emerald, 2009.
[7] H. Schmidt, F. Flerlage, F.J. Hauck, "A generic context service for ubiquitous environments," In Proc. of the IEEE International Conference on Pervasive Computing and Communications (PERCOM), pp.1-6, 2009.
[8] H. Lei, D.M. Sow, J.S. Davis, G. Banavar, and M.R. Ebling, "The design and applications of a context service," SIGMOBILE Mob. Comput. Commun. Rev.*,* vol 6(4), pp.45-55, October 2002.
[9] A. Coronato, G. De Pietro, and M. Esposito, "A Semantic Context Service for Smart Offices," In Proc. of the International Conference on Hybrid Information Technology, vol. 02, pp.391-399, 2006.
[10] C.Bettini, O. Brdiczka, K. Henricksen, J.Indulska, D. Nicklas, A. Ranganathan, D. Riboni, "A Survey of Context Modelling and Reasoning Techniques," Pervasive and Mobile Computing*,* vol. 6(2), pp. 161-180, 2010.
[11] T. Strang, C. Linnhoff-Popien, "A Context Modeling Survey," In Workshop on Advanced Context Modelling, Reasoning and Management, UbiComp 2004 , Nottingham/England, 2004.

AUTHORS PROFILE

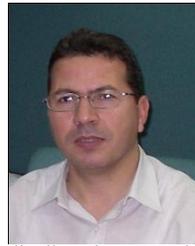

**Elarbi Badidi** is currently an Assistant Professor of computer science at the Faculty of Information Technology (FIT) of United Arab Emirates University. Before joining the FIT, he held the position of bioinformatics group leader at the Biochemistry Department of University of Montréal from 2001 to July 2004. He received a Ph.D. in computer science in 2000 from University of Montréal, Québec (Canada). Dr. Badidi has been conducting research in the areas of object-based distributed systems, bioinformatics tools integration, and Web services. He is a member of the IEEE, IEEE Computer Society, and ACM. He served on the technical program committees of many international conferences. His research interests include Web services and service oriented computing, middleware, cloud computing, and bioinformatics data and tools integration.

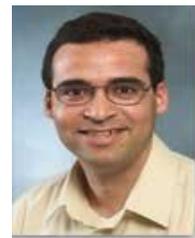

**Larbi Esmahi** is an Associate Professor of the School of Computing and Information Systems at Athabasca University. He was the graduate program coordinator at the same school during 2002-2005. He holds a PhD in electrical engineering from Ecole Polytechnique, University of Montreal. His current research interests are in e-services, e-commerce, multi-agent systems, and intelligent systems. He is an associate editor for the Journal of Computer Science, and the Tamkang Journal of Science and Engineering. He is also member of the editorial advisory board of the Advances in Web-Based Learning Book Series, IGI Global, and member of the international editorial review board the International Journal of Web-Based Learning and Teaching Technologies.